\documentclass{aastex}
\usepackage{graphicx}
\usepackage{txfonts}
\usepackage{subfigure}
\usepackage{natbib}

\begin{document}
   \title{Pair separation of magnetic elements in the quiet Sun}
   \author{F. Giannattasio$^{1}$, F. Berrilli$^{1}$, L. Biferale$^{1}$, D. Del Moro$^{1}$, M. Sbragaglia$^{1}$, L. Bellot Rubio$^{2}$, M. Go\u{s}i\'c$^{2}$, D. Orozco Su\'arez$^{3}$}
   \affil{$^{1}$Dipartimento di Fisica, Università di Roma Tor Vergata and INFN, Via della Ricerca Scientifica,1 I-00133 Rome, Italy\\
   $^{2}$Instituto de Astrof\'isica de Andaluc\'ia (CSIC), Apdo. de Correos 3004, E-18080 Granada, Spain\\
   $^{3}$Instituto de Astrof\'isica de Canarias, E-38205 La Laguna, Tenerife, Spain\\
   \vspace{2cm}
   Final Version (postprint)\\
   Accepted on Astronomy \& Astrophysics 569, A121 (2014)\\
   \vspace{2cm}}
   
   \email{Fabio.Giannattasio@roma2.infn.it}

   \begin{abstract}
    The dynamic properties of the quiet Sun photosphere can be investigated by analyzing the pair 
dispersion of small-scale magnetic fields (i.e., magnetic elements).
    By using $25$ hr-long Hinode magnetograms at high spatial resolution ($0".3$), we tracked $68,490$ magnetic element pairs within a supergranular cell near the disk center.
    The computed pair separation spectrum, calculated on the whole set of particle pairs independently of their initial separation, points out what is known as a super-diffusive regime with spectral index $\gamma=1.55\pm0.05$, in agreement with the most recent literature, but extended to unprecedented spatial and temporal scales (from granular to supergranular). Furthermore, for the first time, we investigated here the spectrum of the mean square displacement of pairs of magnetic elements, depending on their initial separation $r_0$. We found that there is a typical initial distance above (below) which the pair separation is faster (slower) than the average. A possible physical interpretation of such a typical spatial scale is also provided. 
   \end{abstract}

   \shorttitle{Pair separation of magnetic elements in the quiet Sun}
   \shortauthors{F. Giannattasio et al.}

\maketitle

%

\section{Introduction}
\label{Section:Intro}
In the outermost 30\% of the solar radius, the transfer of energy towards the surface occurs via turbulent convection.
To date, a comprehensive theory of solar turbulent convection from small up to global scales has not yet been formulated.
In the last decades, MHD simulations \citep[see, e.g.,][]{1997ASSL..225...79N, 1998ApJ...499..914S, 2001ApJ...546..585S, 2012A&A...539A.121B} have been extensively used to mimic the uppermost convection zone, as they match very well the observations of the solar photosphere. 
However, only tiny regions of the Sun can be realistically simulated, because of the wide range of temporal and spatial convective scales and current computer power.

A complementary approach to investigate the properties of convection on the solar quiet photosphere consists in the study of the interaction between convective flows and the small-scale magnetic fields (hereafter magnetic elements) in the interior of supergranular cells.   
These internetwork magnetic elements can be reasonably regarded as passive objects advected by the underlying flow, as the drag force due to plasma kinetic energy is greater than the magnetic force they exert on the surroundings.
Under this assumption (discussed in Sect.~\ref{Section:Results}), the dynamics of magnetic elements describes that of the plasma \citep[see, e.g., ][]{2011ApJ...727L..30Y, 2013SoPh..282..379B}.
Tracking magnetic elements also allows us to study the onset and amplification of magnetic fields in the quiet Sun, the scales on which they organize, and the rate of interaction between fields \citep[see, e.g., ][]{2012ApJ...752...48C, 2014ApJ...787...87V}. 
This information is important in order to get insights, for example, on the mechanisms that contribute to heating the solar corona, such as magnetic reconnections \citep[see, e.g.,][]{1983ApJ...264..642P, 2006ApJ...652.1734V} and buffeting induced MHD waves \citep[see, e.g.,][]{Stangalini2013, 2013A&A...559A..88S}.

Previous studies have tracked G-band magnetic bright points and magnetic elements from magnetograms (we refer to both of them as magnetic features), regarding them as Lagrangian probes.
Under this assumption, the mean square displacement of such single magnetic features, namely $\langle\Delta l^2\rangle$, has been measured and shown to follow a power law $\langle\Delta l^2(\tau)\rangle\propto\tau^\gamma$, where time $\tau$ is defined as starting from the first detection of the magnetic feature.
In particular, a spectral index $\gamma=1$ is associated with a normal diffusion (also known as random walk) with constant  diffusivity $K\propto\langle\Delta l^2\rangle/\tau$.
In this case, $\langle\Delta l^2\rangle$ corresponds to the standard deviation of a Gauss function describing the distribution of displacements.

It is well known that the presence of the combined effects of the velocity field and a superposed diffusion can lead to a  very large diffusive coefficient,
the so-called eddy-diffusivity,  which is the only relevant parameter needed to predict the long-time, long-space diffusion scale properties in many applied cases \citep{1983Moffatt, Bo90, Cr91, Bi95}. On the other hand, when there is anomalous diffusion (i.e., $\gamma\neq1$) the diffusivity depends on both spatial and temporal scales, and super-diffusive ($\gamma>1$) or sub-diffusive ($\gamma<1$) regimes can arise. The most recent works in the literature agreed that there was a super-diffusive regime in the quiet Sun \citep[see, e.g.,][]{2011ApJ...743..133A, 2013ApJ...770L..36G, 2014ApJ...788..137G, 2014A&A...563A.101J}.  
This implies that the effective diffusivity decreases when the temporal (spatial) scale is reduced \citep{2011ApJ...743..133A}, thus allowing magnetic fields to be enhanced on the very short (small) scales.

Diffusive (normal or anomalous) regimes are typically well defined only in the asymptotic limit of large time and large distance, something that is very difficult to achieve in our case. 
In many geophysical and astrophysical situations only the transient behavior is observable and/or relevant, hence the interest in discussing in a more quantitative way the separation of pairs of magnetic elements in our observational set-up.
While the diffusion of single Lagrangian probes is dominated by large scale motions, pair separation should be more universal, being  affected only by the relative velocity fields on scales on the order of the  distance between the magnetic elements.
If we identify with  $r_\tau = |\mathbf{X}^{(1)}_\tau-\mathbf{X}^{(2)}_\tau| $ the distance between two magnetic elements in our ensemble at a given time $\tau$, we will be interested in quantifying the probability distribution function (PDF), $p(r,\tau|r_0, \tau_0)$,  of observing a given separation $r$ at time $\tau$ starting from an initial distance $r_0$ at time $\tau_0$.  In general, we obtain 
\begin{equation}
\label{eq:diff1}
\langle\Delta r^2\rangle\equiv\langle |r_\tau-r_0|^2\rangle= \int_0^\tau  \langle \delta_{r_{t}} v \delta_{r_0} v \rangle dt, 
\end{equation}
where $\delta_r v $ is the velocity difference among the magnetic elements, the average is meant over all the considered pairs in our ensemble (see Sect.~\ref{Section:Observations}) and we have assumed that the initial position is uncorrelated from the underlying velocity field \citep{So99}.  In many cases, the integral in Eq.~\ref{eq:diff1} is well behaved and in the limit of large time converges to $\propto \langle v^2\rangle \tau$, i.e., we have asymptotically a normal diffusion process with diffusivity given by the one-point velocity fluctuations along the trajectories of the magnetic elements. Nevertheless, there are many cases where this asymptotic regime is never reached, or it lies just outside the spatial and temporal limits of observation. For example, it is well  known that in the presence of multi-scale non trivial statistical properties for the underlying velocity field, anomalous diffusion might develop.
This is for instance the case of the celebrated Richardson diffusion \citep{Richardson} for homogeneous and isotropic turbulent flows in the range of spatial scales where the velocity obeys a Kolmogorov 1941 turbulent cascade \citep{Fr95}. In this case it is predicted and observed the presence of a super-diffusion with  $\gamma=3$ and self-similar Richardson-like PDF  \citep{Ju99, Fa01, Ye04, Mo07, Sa09, Be10}. 
Similarly, it is known that in the presence  of a spatially smooth velocity field with very long temporal correlation, there might be an anomalous (super- or sub-) diffusion \citep[see, e.g., the pioneering works of ][]{Ge85, Av92, Za93}. 
Finally, there might also be the possibility of observing {\it strong} anomalous diffusion, i.e., a PDF for pair separation that is not self-similar \citep{Ca99}, something that has been recently detected in turbulent flows using a very high-statistical dataset \citep{Sc12}. 
In all cases, we are in the presence of a sort of strong or weak failure of the central limit theorem, i.e., the right-hand side of Eq.~\ref{eq:diff1} cannot be trated as the sum of many uncorrelated variables, either because we are exploring spatial and temporal scales too small compared with the characteristic variations of the underlying velocity, or because the velocity field itself possesses non-trivial spatial and temporal asymptotic multi-scale properties. 

A {\it  local} (in temporal and spatial scale) effective diffusion coefficient can be defined as
\begin{equation}
\frac{d}{d\tau} \langle\Delta r^2\rangle=K(r_0,\tau, r),
\end{equation}
which gives us the typical separation speed of two magnetic elements found at separation $r$ after a time $\tau$ and with initial separation $r_0$.
When the central limit theorem holds, we have $K \sim const.$, independently of the original separation $r_0$. 
Otherwise, different anomalous regimes arise. 
In particular, here we are interested in studying the effects of $r_0$ on pair separation for the case of magnetic elements in the quiet Sun, where the observational and intrinsic physical limitations do not allow us to extend the observing regimes to couples with separations much larger than $r_0$, and therefore where only the pre-asymptotic regime is observable.  
For instance, for fully developed turbulent flows, it is known that the pair separation is only ballistic, namely $\langle\Delta r^2\rangle \sim t^2$, up to a maximum time, the so-called Batchelor time \citep[i.e., $t_B$, ][]{Bo06},  wich depends on the initial separation and follows the law $t_B \propto r_0^{2/3}$. 

To our knowledge, only \citet{1998ApJ...506..439B} and \citet{2012ApJ...759L..17L} have applied pair dispersion analysis to magnetic features in the photosphere.
\citet{1998ApJ...506..439B} tracked $622$ G-band bright point pairs acquired at the SVST by observing a region $29"\times29"$ wide for $70$ minutes, with a cadence of $25$ s and a spatial resolution of $\sim0".2$. 
They measured a spectral index consistent with $\gamma\simeq1.3$.
\citet{2012ApJ...759L..17L} used NST images \citep{2010AN....331..620G} to track a maximum of $7912$ magnetic bright points in a quiet Sun region, a coronal hole, and a plage region.  
They measured a spectral index $\gamma\simeq1.5$ everywhere in the temporal range $10\lesssim\tau\lesssim400$ s.
They interpreted this difference from known scaling as due to the imperfectly passive nature of bright points.

The limitation in the statistics prevented the aforementioned authors from carrying out pair dispersion analysis taking into account the initial separation $r_0$.
In this work, for the first time the pair separation approach is used to investigate the dynamic behavior of magnetic elements in the quiet Sun for different initial pair separations. 
We believe this is an important point, since it is difficult to imagine that pair separation is not affected by the initial condition for those temporal and spatial scales accessible on the Sun. 
We present the results obtained with $68,490$ tracked magnetic pairs.

\section{Observations and data analysis}
\label{Section:Observations}
\begin{figure}[ht!]
  \centering
    \resizebox{\hsize}{!}{\includegraphics {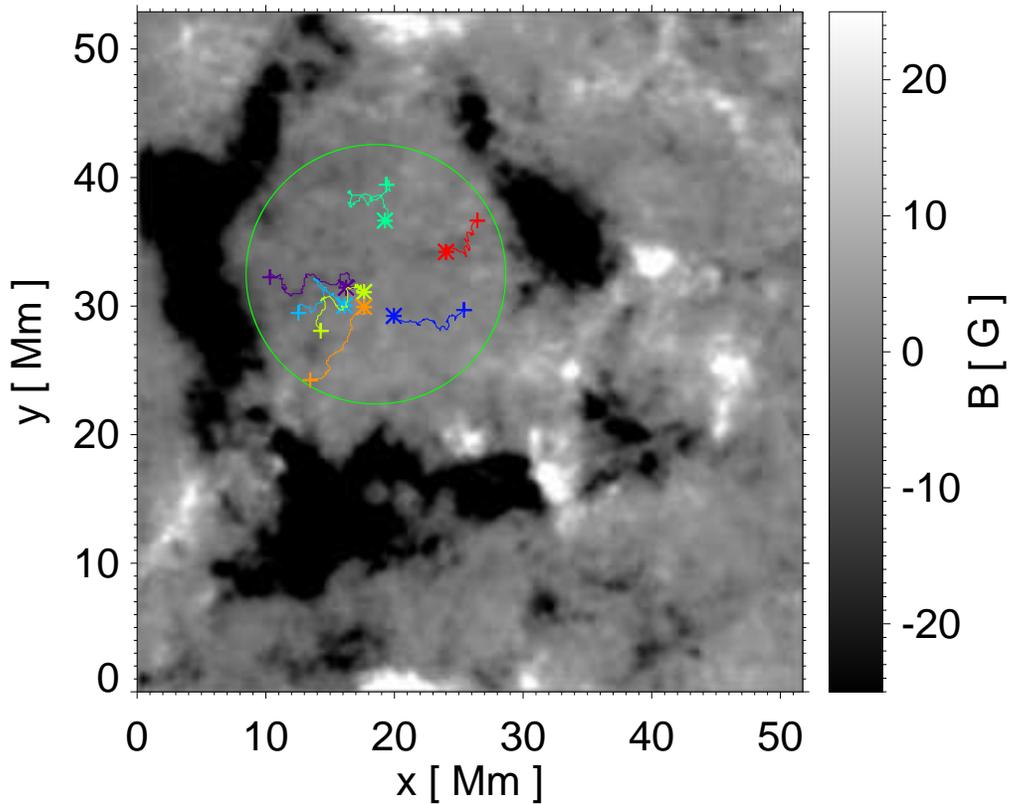}}
    \caption{Time-averaged magnetogram saturated at $25$ G. Only the magnetic elements inside the ROI (limited by the green circle) are considered for the analysis. A few trajectories of magnetic elements are shown in the ROI. The asterisks mark the first detection positions; the plus signs mark the last detection positions.} 
\label{Fig:Mag}
\end{figure}
\begin{figure*}[ht!]
  \centering
    \includegraphics{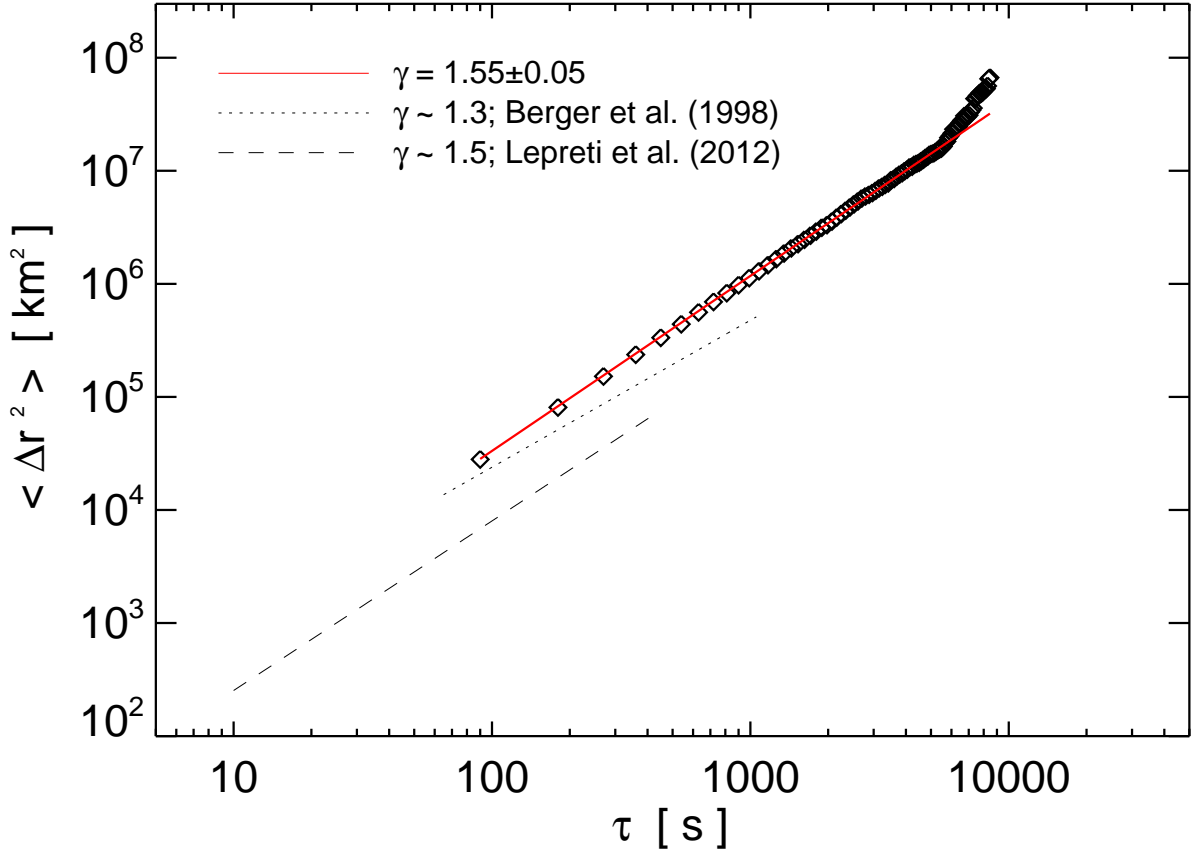}
    \caption{Mean square pair separation as a function of time since the first appearance. 
    Only the data points up to $\sim5500$ s have been used to make the fit. 
    The error on $\gamma$ was computed as the standard deviation of the values obtained after a random subsampling of magnetic pairs.
    The results described in \citet{1998ApJ...506..439B} (dotted line) and \citet{2012ApJ...759L..17L} (dashed line) are superposed for comparison.}  
\label{Fig:SepSpectrum}
\end{figure*}
\begin{figure}[ht!]
  \centering
    \resizebox{\hsize}{!}{\includegraphics {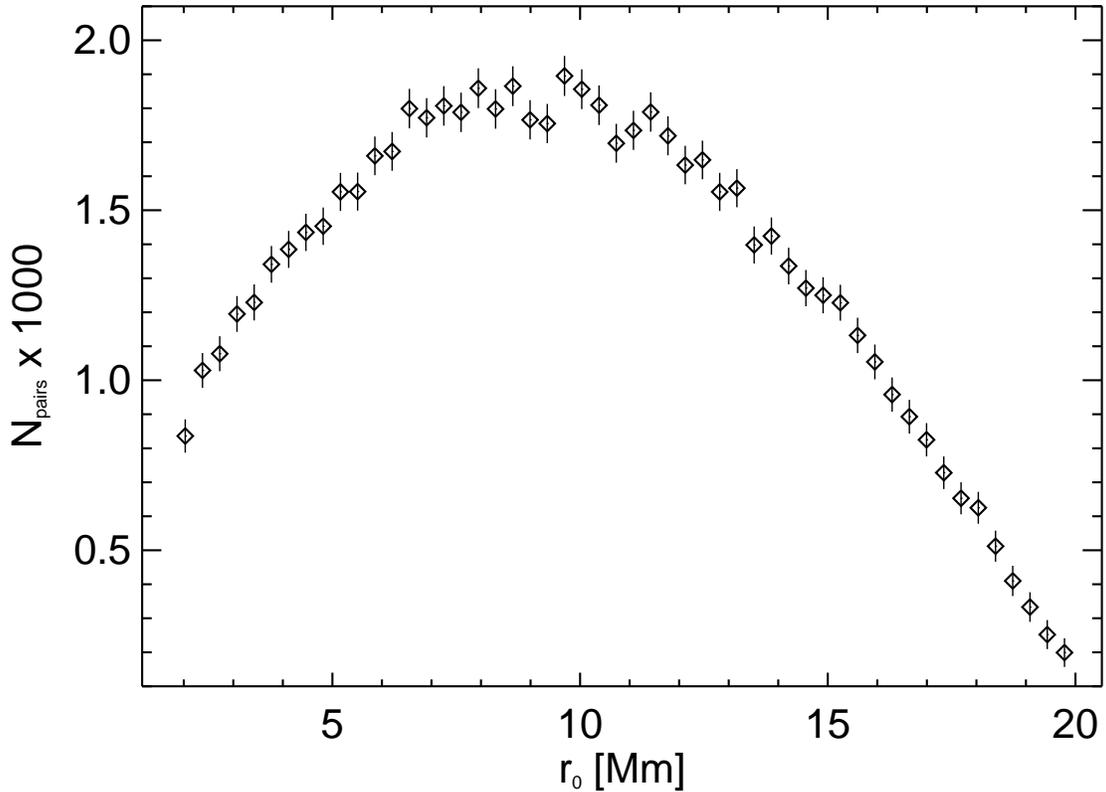}}
    \caption{
    Number of magnetic pairs as a function of the initial separation. Vertical bars represent errors (see the text).
    } 
\label{Fig:InitialSep}
\end{figure}
\begin{figure*}[ht]
   \centering
   \resizebox{\hsize}{!}{\includegraphics {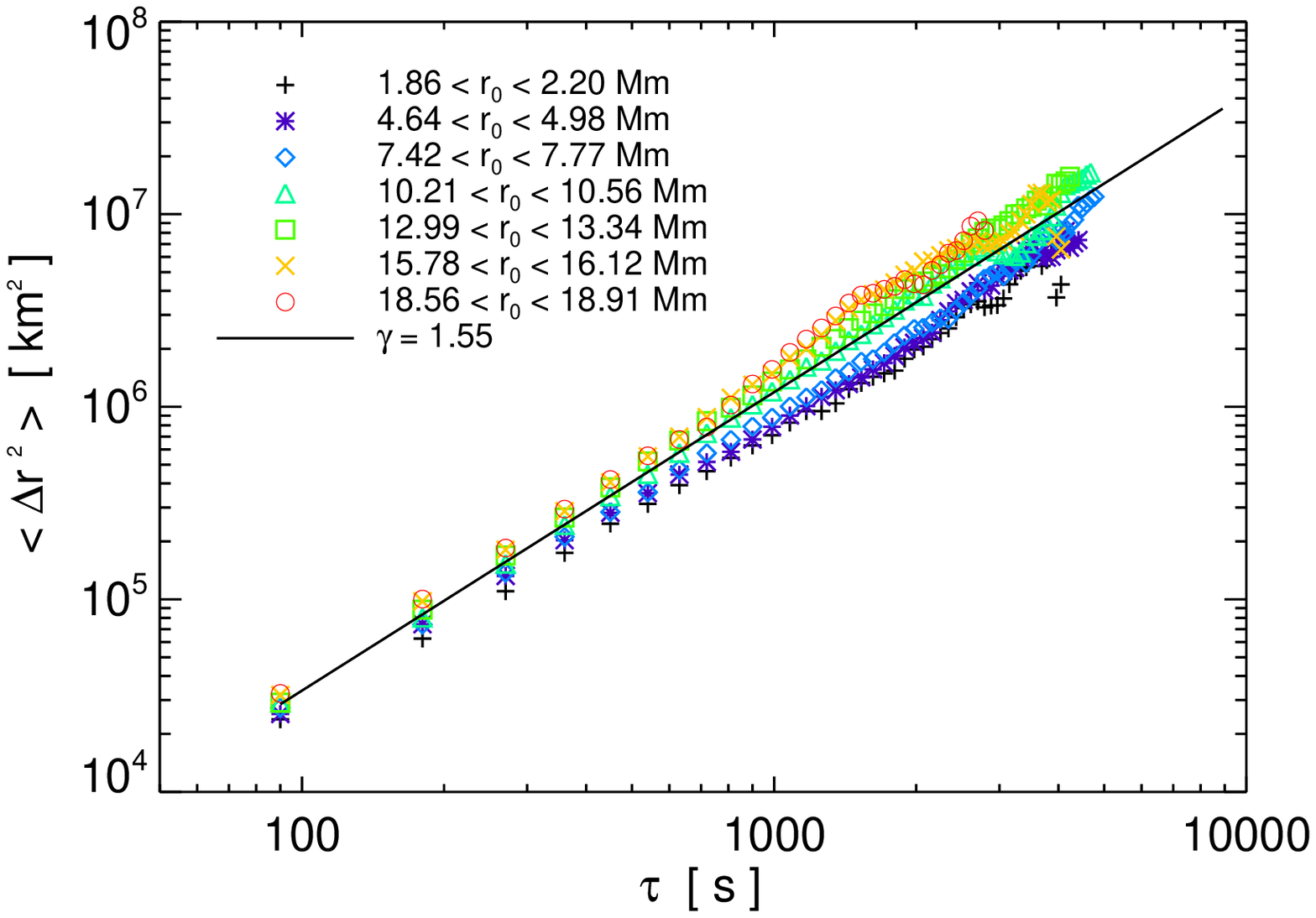}}
   \begin{picture}(0,0)
    \put(40,85){\includegraphics[height=4.cm]{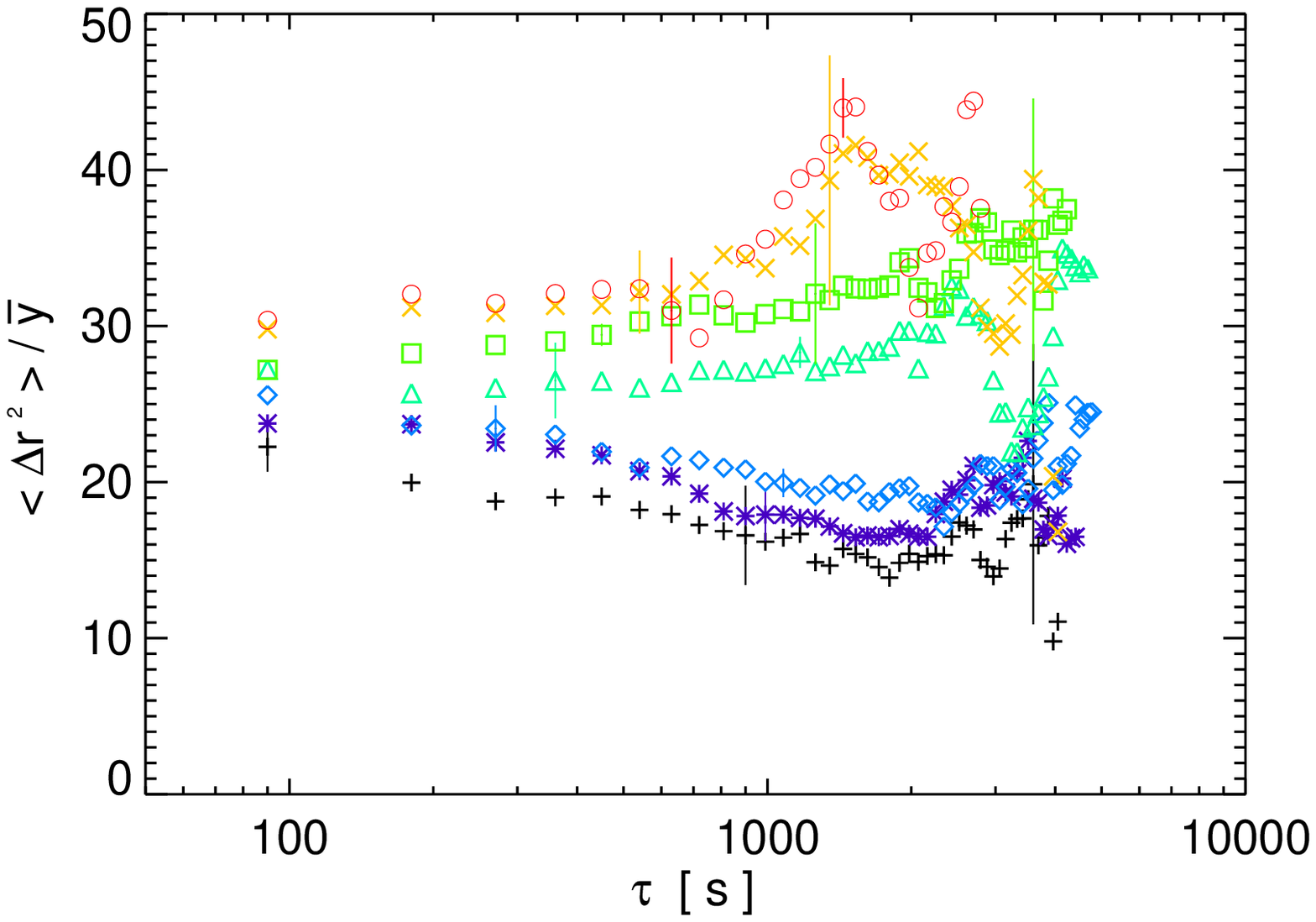}}
   \end{picture}

   \caption{Mean square separation $\langle\Delta r^2(\tau,r_0)\rangle$ for seven different and equally spaced values of $r_0$. The black solid line corresponds to the fitting curve $\bar{y}$ of Figure \ref{Fig:SepSpectrum}. In the inset the compensated mean square separation $\langle\Delta r^2(\tau,r_0)\rangle/\bar{y}$ is shown. The errors (vertical bars) are shown only for a few data points.}
   \label{Fig:SepR0}
\end{figure*}
\begin{figure*}[ht]
   \centering
   \subfigure[]{\includegraphics [width=7cm]{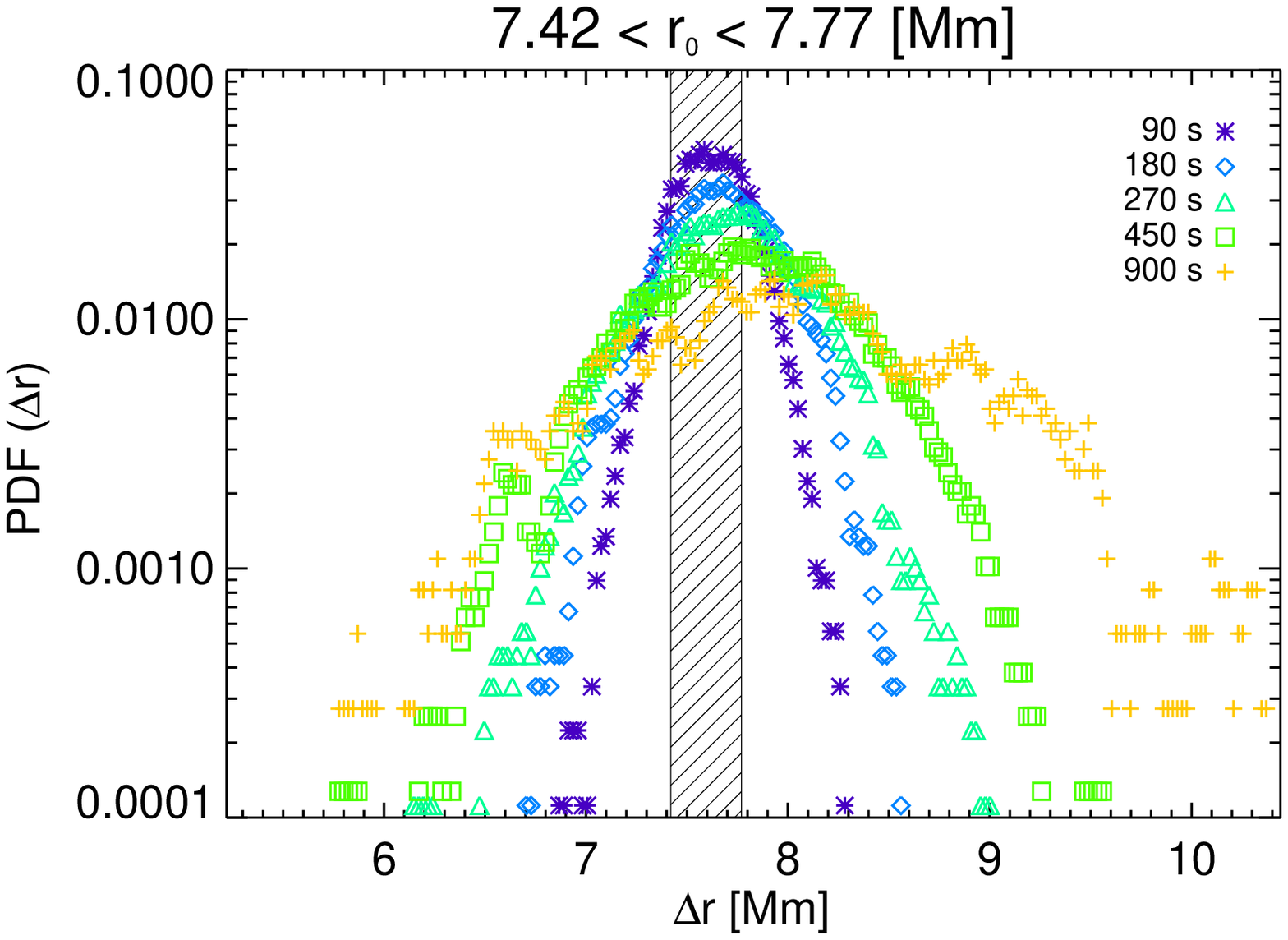}}
   \subfigure[]{\includegraphics [width=7cm]{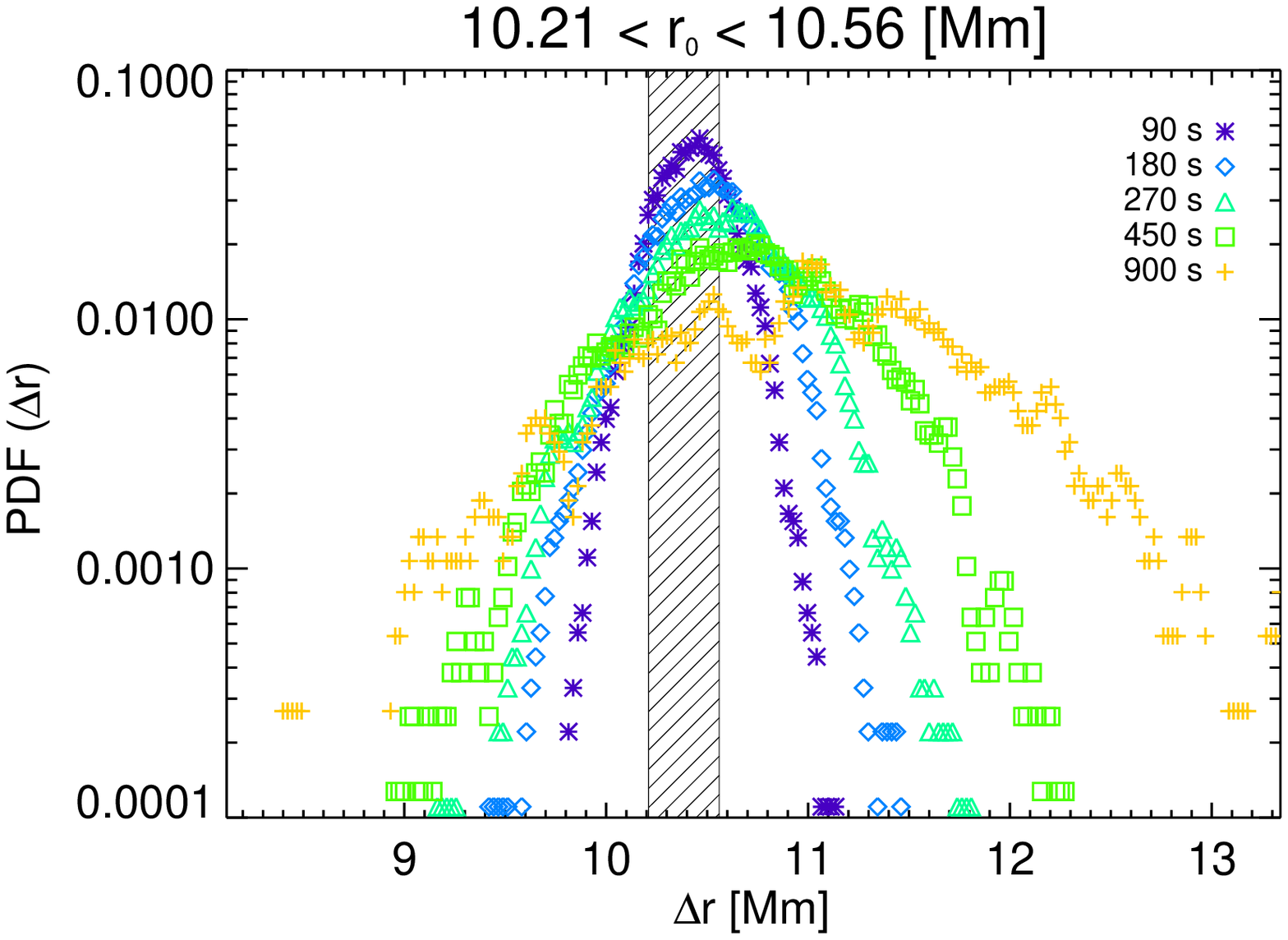}}
   \subfigure[]{\includegraphics [width=7cm]{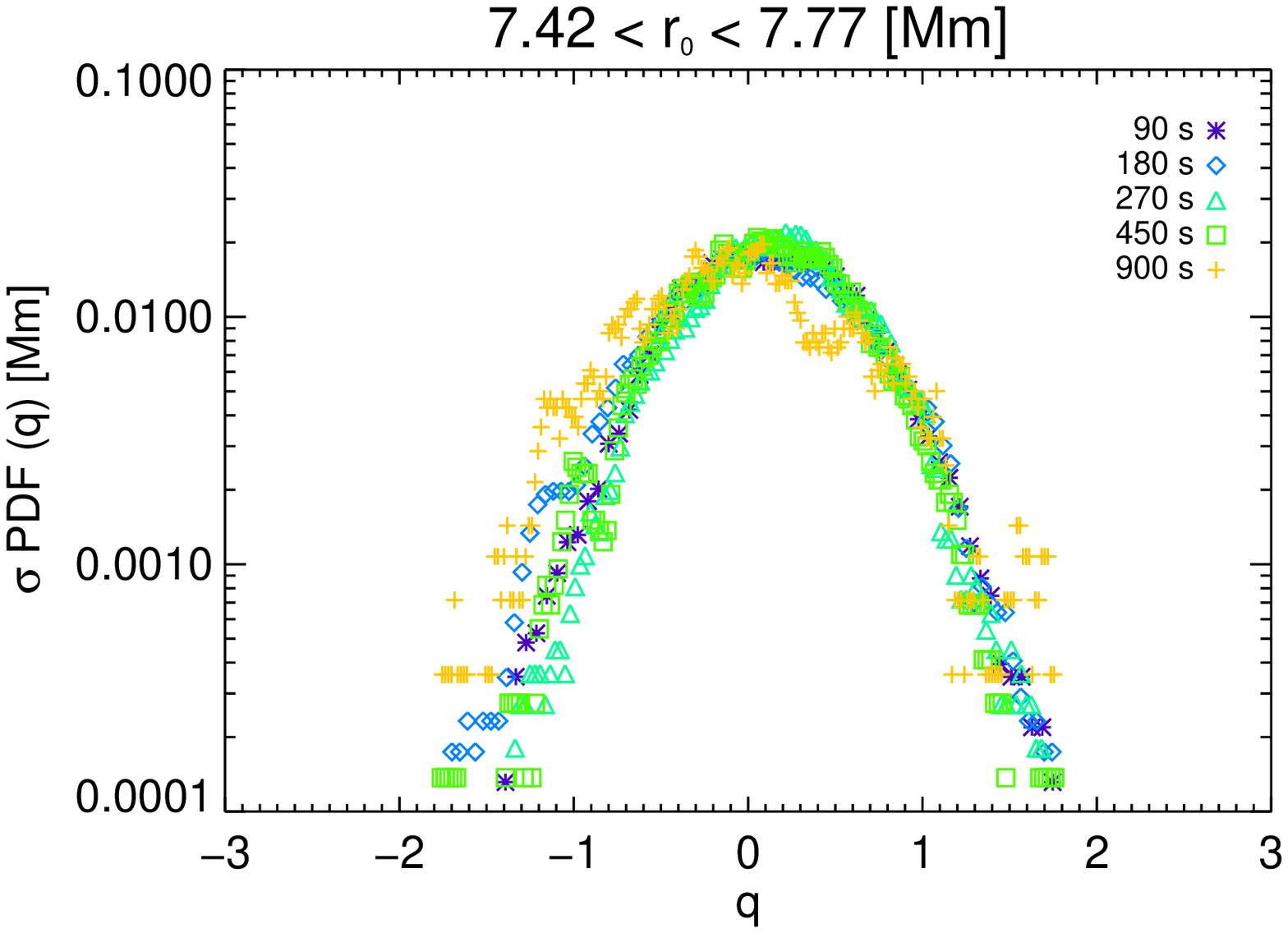}}
   \subfigure[]{\includegraphics [width=7cm]{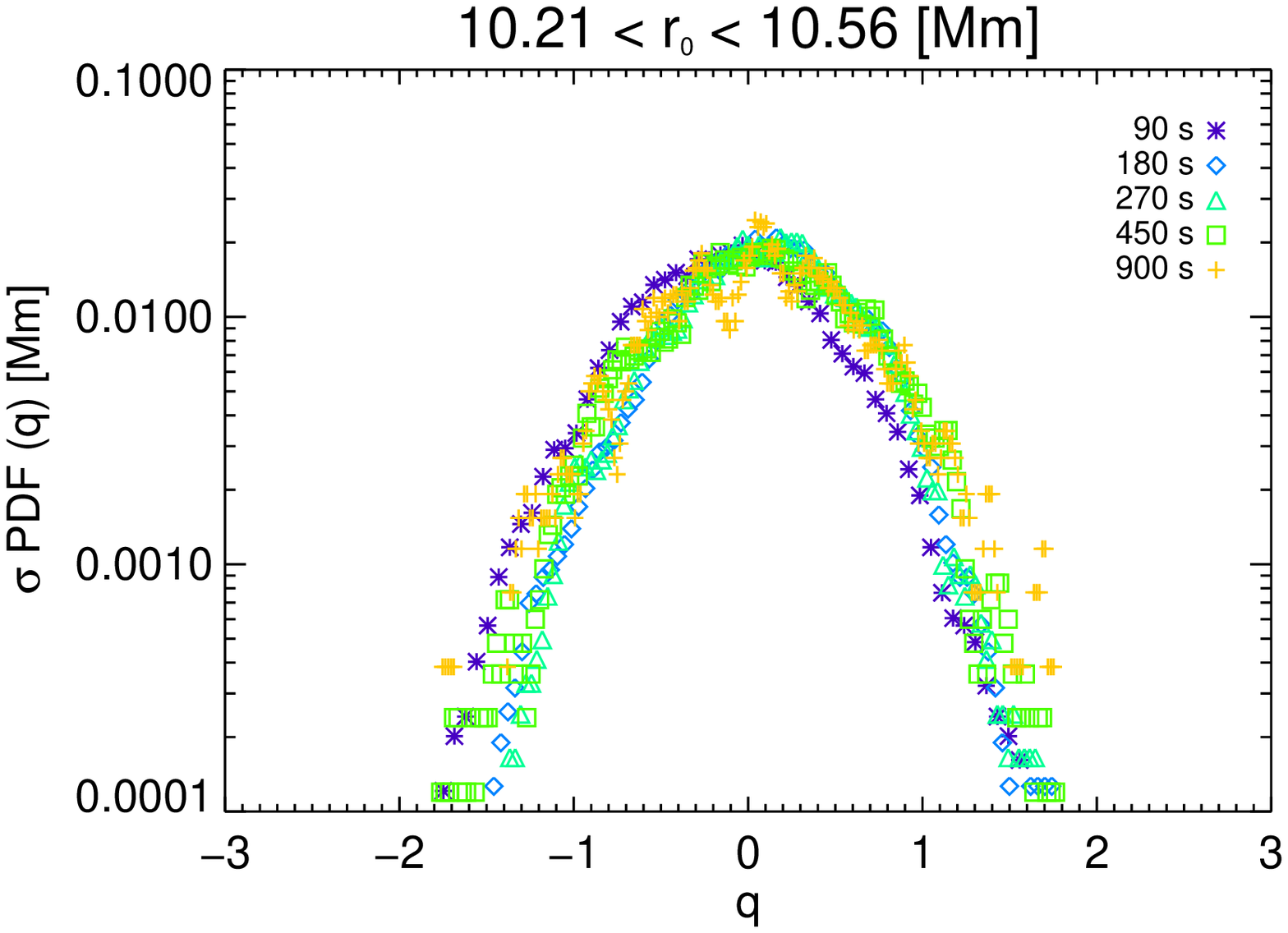}}
   \caption{Time-dependent PDF for pair separation of magnetic elements (a, b), and the same rescaled to unity rms (c, d) for initial separations in the range $7.42<r_0<7.77$ Mm (a, c) and $10.21<r_0<10.56$ Mm (b, d). The shaded areas in panels (a) and (b) cover the initial separation bins. All PDFs are normalized to unit area.} 
   \label{Fig:PDFs}
\end{figure*}
\noindent
The data set used in this work was described in \citet{Milan} and analyzed by \citet{2013ApJ...770L..36G, 2014ApJ...788..137G} to study the diffusion of single magnetic elements up to supergranular scales.
It consists of $995$ Hinode-NFI magnetograms \citep{2007SoPh..243....3K, 2008SoPh..249..167T} with a spatial resolution of $0".3$ and a noise of $\sigma_B=6$ G for single frames. 
The magnetograms were co-aligned, trimmed to the same field of view (FoV, which is $\sim50$ Mm sized), and filtered out for oscillations at $3.3$ mHz \citep{Milan}.
As a consequence, the present analysis is free from effects of acoustic oscillations and atmospheric seeing, and aimed to magnetic elements and not magnetic proxies like G-band bright points.
The large FoV, which encloses an entire supergranule, and the high spatial resolution enabled us to investigate a wide range of spatial scales and observe a large number of magnetic elements.
The series, acquired on November 2, 2010, covers 25 hours without interruption, with a cadence of $90$ s.
This allowed us to study the dynamics of magnetic elements on a wide range of temporal scales (from a minute to a day).

In Figure \ref{Fig:Mag}, we show the $25$ hr time-averaged magnetogram of the FoV saturated at $25$ G. 
We focus on the region of interest (ROI) inside the green circle, which is centered in the center of the supergranule and has a radius of $\sim10$ Mm, such that the ROI is completely enclosed within the supergranule itself, i.e., in the internetwork region \citep{2014ApJ...788..137G}.

We applied the tracking algorithm described in \citet{2004A&A...428.1007D}.
The algorithm uses a variable threshold in order to overcome the loss of weak fields and resolve the clustered peaks of the largest magnetic features \citep[e.g., ][]{2005SoPh..228...81B}.
We discarded all the magnetic elements with speed $v>7$ kms$^{-1}$, which is roughly the speed of sound in the photosphere.
We also discarded the magnetic elements passing close to the boundary of the ROI (at a distance $\lesssim1.86$ Mm).
A total of $68,490$ tracked magnetic pairs originated in the ROI have been detected.
In Figure \ref{Fig:Mag} we also show, for the sake of visualization, the evolution of a few magnetic elements forming a subset of pairs.

\section{Results and discussion}
\label{Section:Results}

\noindent
The detected pairs have been used to investigate the nature of turbulent convection under the hypothesis of magnetic elements passively transported by the supergranular flows in the ROI.
By analyzing the same data set, \citet{2013ApJ...770L..36G} found that the equipartition magnetic flux density is $B_e\simeq255$ G. 
Only less than $4\%$ of the total magnetic elements have an average flux greater than that value, and are all located in the network.
Therefore, the condition of passive magnetic elements is reasonably fulfilled in the internetwork regions, as demonstrated by the spectro-polarimetric studies performed by \citet{2007ApJ...670L..61O, 2012ApJ...751....2O}; and \citet{2012ApJ...757...19B}.

As done in the previous works in the literature \citep[see, e.g.,][]{2012ApJ...759L..17L} the separation spectrum $\langle\Delta r^2\rangle$ was first computed for all the pairs of magnetic elements, regardless of their initial separation.
We obtained the results shown in Figure \ref{Fig:SepSpectrum} (black diamonds). 
We found that the separation spectrum is best fitted by a power law $\bar{y}\propto\tau^\gamma$ with spectral index $\gamma=1.55\pm0.05$ (the red line in the same figure), in agreement with the results of \citet{2012ApJ...759L..17L}, but here extended to supergranular scales.
The error on $\gamma$ was computed as the standard deviation of the values obtained after a random subsampling of the magnetic pairs.

As mentioned in Sect.~\ref{Section:Intro}, the high number of magnetic elements tracked allowed us to perform for the first time the pair separation analysis for different values of the initial pair separation.
By looking in greater detail at the distribution of initial separations, one easily recognizes that it is in general very broad. 
In Figure \ref{Fig:InitialSep} we show, for each initial separation $r_0$, the number of magnetic pairs found in the range $r_0-0.174<\Delta r<r_0+0.174$ Mm, which is $3$ Hinode-NFI pixels wide.
Vertical bars in the graph represent the errors, which are mainly due to the Poissonian contribution.
The peak of the curve lies between $8$ Mm and $9$ Mm ($r_{0,peak}$).
By comparing Figs. \ref{Fig:SepSpectrum} and \ref{Fig:InitialSep} one can see that the global mean displacement with respect to the initial separation, averaged over all pairs, is of the same order of magnitude as the  spread in the initial distribution of $r_0$ in our sample, i.e., we have not reached any asymptotic long-time regime. Hence, it is natural to ask the question: How robust is the observed super-diffusive behavior as a function of $r_0$?
As magnetic elements are not point-like, but have diameters up to $d_{min}\simeq1.86$ Mm, it is not possible to chose arbitrarily small mutual distances. 
Therefore, the minimum pair separation set is $d_{min}$. 
Moreover, the maximum achievable separation is given by the diameter of the ROI, which is $d_{ROI}\sim20$ Mm. 
Thus, $r_0$ must satisfy $d_{min}\le r_0\le d_{ROI}$. 
We choose bins of $r_0$ large enough to collect a sufficiently high number of magnetic pairs within, and small enough to be able to study the variation of $\langle\Delta r^2\rangle$ with $r_0$.
For this purpose, we set the bin size at $348$ km (i.e., $3$ Hinode-NFI pixels), and computed $\langle\Delta r^2(\tau,r_0)\rangle$ for each bin.
In Figure \ref{Fig:SepR0} we plot seven of all the computed $\langle\Delta r^2(\tau,r_0)\rangle$.
For comparison, we also over-plot the power law behavior $\bar{y}(\tau)$ (corresponding to $\gamma=1.55$).
In order to emphasize the deviations from such a law, in the inset of the same figure we plot a compensated pair separation spectrum $\langle\Delta r^2(\tau,r_0)\rangle/\bar{y}$.
The errors on data points were computed as the standard deviation of the values obtained after a random subsampling of the magnetic pairs.
From these two plots we can deduce that
1) the smaller the $r_0$, the smaller the effective eddy diffusivity  $\langle\Delta r^2\rangle/\tau$ for any $\tau$, and 2) the smaller the $r_0$, the smaller the value of the {\it effective} slope $\gamma$; in addition, there is a clear change in the trend for initial separation crossing the value $r_0\sim10$ Mm, which roughly corresponds to the radius of the ROI.
From an observational point of view, 1) and 2) could be interpreted by taking into account the recent results in \citet{2012ApJ...758L..38O} and \citet{2014ApJ...788..137G}.
In those works, the authors showed that the horizontal velocity field within a supergranule is mostly radial and directed from the center to the boundaries.
Following this sketch, we expect that magnetic elements starting close to each other will, on average, separate more slowly than magnetic elements starting farther away from one another. 
In fact, magnetic elements with a larger initial separation are most likely to be dragged on along very different directions, thus separating faster. 
This effect naturally introduces a dependence of $\langle\Delta r^2\rangle$ on $r_0$.
In particular, the systematic  increase of the effective  $\gamma$ from the granular to the supergranular scale suggests that the pre-asymptotic 
diffusion is a function of the probed spatial scale.

By comparing figures \ref{Fig:InitialSep} and \ref{Fig:SepR0} we note that for initial separations smaller than $r_{0,peak}$ the slopes in the pair separation spectrum decrease at longer times; while for larger initial separations the slopes increase.
This trend is significant, as can be seen from the errors on the data points shown in the inset in Figure \ref{Fig:SepR0}. 
The pairs of magnetic elements with initial separation around $r_{0,peak}$ (from $\sim6.5$ to $\sim11.5$ Mm), which are in number about half of the entire population of pairs, are characterized by a separation spectrum with $\gamma$ around $\simeq1.55$.
This explains why in the separation spectrum in Figure \ref{Fig:SepSpectrum}, which was retrieved by considering all the pairs of magnetic elements in the ROI (with any initial separation), there is an effective trend consistent with $\gamma\simeq1.55$ even at longer temporal scales. 

To further investigate the effects of $r_0$ on the pair separation, we computed the time-dependent PDF (normalized at unit area) of observing a given separation starting from the initial values of $7.42<r_0<7.77$ Mm and $10.21<r_0<10.56$ Mm, at which values the change in the trend shown in Figure \ref{Fig:SepR0} is observed.
The results are shown in panels (a) and (b) of Figure \ref{Fig:PDFs}.
In that figure, the initial separation range is depicted as a shaded area.
As we can see, the PDF broadens with time, its rms being $\sigma(\tau)$, and the peak moves to gradually increasing separations.
At $\tau=900$ s the tails begin to be important, and affect substantially the pair separation spectrum, indicating that the largest separations begin to become dominant.
We rescaled the time-dependent PDF so that it is centered at zero, and its rms $\sigma(\tau)$ is unity \citep[see, e.g.,][]{Ju99}.
To this end, we introduced the rescaled separation $q=(\Delta r-\langle\Delta r\rangle)/\sigma$, being $\langle\Delta r\rangle$ the mean separation value, and computed PDF($q$) at each time.
The correct re-normalization required to consider $\sigma$PDF(q) instead of PDF(q).
The results are shown in panels (c) and (d) of Figure \ref{Fig:PDFs}.
As we can see, the curves seem to collapse on each other, especially in proximity of $\langle\Delta r\rangle$ (which corresponds to $q=0$).
This means that the small deviations from the mean values seem to be {\it self-similar}.
When it holds, self-similarity indicates that there is a single underlying distribution governing the process at any time \citep{Ju99}. 
However, in our case the statistics is still too low to come to a conclusion in this sense.
More data points are nedeed to better sample the tails of the time-dependent PDF, where any possible breaking of self-similarity can be detected.



\section{Conclusions} 
\label{Section:Conclusions}
Dynamic processes in the solar photosphere can be studied at spatial and temporal scales from granular to supergranular by measuring the pair separation rate of quiet Sun magnetic elements.
By taking advantage of uninterrupted $25$ hr magnetograms acquired by Hinode at high resolution and imaging a whole supergranule, we computed the separation spectrum of $68,490$ pairs of magnetic elements.
When considering all the pairs within the supergranule, we found a spectral index $\gamma=1.55\pm0.05$, in agreement with the most recent literature, but extended at unprecedented spatial and temporal scales.
Such a super-diffusive regime can be interpreted as being due to an underlying velocity field with either characteristic spatial (temporal) scales larger (longer) than the scales of observation; or non-trivial asymptotic multi-scale properties.
For the first time we investigated the separation spectrum for different values of the initial pair separation of magnetic elements, $r_0$.
The main conclusion is that the rate of pair separation depends on the spatial scale under consideration.

The possibility that the pre-asymptotic diffusive behavior detected here possesses non-trivial multi-scaling properties remains to be investigated; in other words, whether higher order moments do not scale proportionally to the second order moment, $\langle (\Delta r)^{2p} \rangle \neq  \langle (\Delta r)^{2} \rangle^p $. 
This would indicate the presence of {\it strong} anomalous diffusion \citep{Ca99}, possibly connected to the presence of intermittent properties of the advecting velocity field. This ambitious goal surely represents a great challenge for future research since it can only be achieved by extending by at least one order of magnitude the statistical ensemble and the temporal window of the observation.
\begin{acknowledgements}
 This work was supported by a PhD grant at the University of Rome, Tor Vergata.
 Part of this work was done while F.G. was a Visiting Scientist at Instituto de Astrof\'isica de Andaluc\'ia (CSIC). Financial support by the Spanish MEC through project  AYA2012-39636-C06-05 (including European FEDER funds) is gratefully acknowledged.
 This work has also benefited from discussions in the Flux Emergence meetings held at ISSI, Bern in December 2011 and June 2012.
 The data used here were acquired in the framework of {\itshape Hinode} Operation Plan 151, entitled {\itshape Flux replacement in the solar network and internetwork}.\\
 L.B. acknowledges partial funding from the European Research Council under the European Community’s Seventh Framework Programme, ERC Grant Agreement No 339032.\\
 {\itshape Hinode} is a Japanese mission developed and launched by ISAS/JAXA, collaborating with NAOJ as a domestic partner, NASA and STFC (UK) as international partners. Scientific operation of the Hinode mission is conducted by the Hinode science team organized at ISAS/JAXA. This team mainly consists of scientists from institutes in the partner countries. Support for the post-launch operation is provided by JAXA and NAOJ (Japan), STFC (U.K.), NASA, ESA, and NSC (Norway).
 \end{acknowledgements}

\end{document}